\documentclass[twocolumn,twoside,slac_two]{revtex4}
\usepackage{graphicx}
\usepackage{fancyhdr}
\newcommand{\gtae}{\buildrel {\lower3pt\hbox{$>$}} \over 
{\lower2pt\hbox{$\sim$}} }
\newcommand{\ltae}{\buildrel {\lower3pt\hbox{$<$}} \over
{\lower2pt\hbox{$\sim$}}}

\begin{document}

\title{Runaway core collapse and cluster survival: where 
are the parent clusters of ULXs?}

%

\author{Roberto Soria}
\affiliation{Harvard-Smithsonian Center for Astrophysics, 
Cambridge, MA 02138, USA}
%

\begin{abstract}
Accreting intermediate-mass black holes (IMBHs) have been proposed  
as an explanation for ultraluminous X-ray sources (ULXs).
Runaway core collapse inside a massive cluster 
is a possible mechanism for IMBH formation. 
But if so, why are ULXs only rarely found associated 
with a cluster? We use a simple analytical approximation 
to show that rapid core collapse can occur 
in two physical regimes. For cluster masses $\sim 10^6 M_{\odot}$, 
an IMBH may be formed if the collapse timescale is $\ltae 3$ Myr, 
as already well known; the parent cluster is expected 
to survive. For cluster masses $\sim 10^5 M_{\odot}$, an IMBH may result 
from a core collapse on even shorter 
timescales ($\approx 0.5$ Myr), when the protocluster 
is still embedded in optically thick gas. Most clusters 
in this latter regime are disrupted ``explosively'' as soon as  
the gas is ionized by the OB stars. We speculate 
that this process may produce isolated ULXs 
with masses up to a few $10^2 M_{\odot}$, surrounded by 
a loose OB association, and perhaps by a nebula, remnant 
of the expanding gas from the disrupted protocluster. 

\end{abstract}

\maketitle

\thispagestyle{fancy}


\section{INTRODUCTION}

Various alternative models have been proposed
to explain the nature of ULXs
in nearby galaxies. Inhomogeneous accretion flows
may allow luminosities a factor of $10$ 
in excess of the Eddington limit~(\cite{Begelman}). 
Moderate geometrical beaming (\cite{Kingetal2001}) 
in the direction of the observer
may boost the observed brightness by an order
of magnitude. Thus, sources with an apparent
isotropic X-ray luminosity up to a few $10^{39}$
erg s$^{-1}$ can be explained by stellar-mass
black holes (BHs) with masses $\approx 5$--$15$ M$_{\odot}$,
without the need to invoke new astrophysical mechanisms.
However, these scenarios become problematic
for ULXs with steady apparent X-ray luminosities
$> 10^{40}$ erg s$^{-1}$: they require either
strong beaming, such as a relativistic jet
pointing towards the observer (\cite{FabrikaMescheryakov2001}; 
\cite{KordingFalckeMarkoff2002}),
or a more massive accretor, with
$M \sim 100$--$1000$ M$_{\odot}$ (IMBHs: 
\cite{ColbertMushotzky1999}).
It is possible that ULXs are a mixture of different
classes of objects. Observational and statistical
arguments have been made against a predominance
of beamed sources (\cite{DavisMushotzky2004}).
In other cases, circumstantial evidence in favor
of IMBHs has been proposed from X-ray spectral
and timing studies~(\cite{MillerFabianMiller2004}).

One of the difficulties of the IMBH scenario
is how to introduce a plausible formation mechanism.
It has been suggested that IMBHs could
be the remnants of very massive ($\sim 1000$ M$_{\odot}$)
zero-metallicity Population-III stars (\cite{MadauRees2001}).
However, this is difficult to reconcile
with the relative abundance of ULXs in young
star-forming regions, especially in starburst
and merging galaxies.
Another possibility (\cite{KingDehnen2005})
is that at least some ULXs may be the nuclei
of smaller satellite galaxies
accreted and tidally disrupted by a larger host
(\cite{BekkiFreeman2003}).

Alternatively, it has been proposed (\cite{Gurkanetal2004}, 
\cite{PzMc2002}, \cite{Pzetal2004}) 
that IMBHs could be formed in the dense cores of massive,
young stellar clusters or super star clusters.
Analytical approximations and numerical simulations 
have shown that, for a suitable range of cluster
masses and densities, mass segregation and the Spitzer
instability (\cite{Spitzer1987}) can lead to a runaway core collapse
on a timescale shorter than the lifetime of
its most massive stars ($\approx 3$ Myr).
The same simulations also show that
the mass of the collapsing core is $\sim 10^{-3}$
times the total mass of the cluster. Thus,
this scenario could explain the formation of
IMBHs with a mass up to $\sim 1000$ M$_{\odot}$.

The main difficulty of theoretical models invoking
IMBH formation inside a cluster is that, in fact,
most ULXs do not have a bright, massive cluster
as their optical counterpart, apart for few notable
exceptions (e.g., a bright ULX in M\,82 is associated with
the star cluster MGG-11: \cite{Pzetal2004}).
In some cases, such as the ULXs in the Cartwheel Galaxy
(\cite{Gaoetal2003}), the distance is such that
we cannot draw firm conclusions on the significance
of possible ULX-cluster coincidences.
In the starbursting Antennae Galaxies, most ULXs are
displaced by $\approx 300$ pc from their nearest
star clusters (\cite{ZezasFabbiano2002}, \cite{Zezasetal2002}).
In the starburst galaxy NGC\,7714, no optical
counterparts have been found for the
two brightest ULXs (\cite{SoriaMotch2004}, \cite{Smithetal2005}):
there are no objects brighter than $M_V \sim M_B \sim -9$
within a few hundred pc of the two sources.
This rules out young clusters with masses
$\gtae 10^4$ M$_{\odot}$. In other cases (e.g., 
in NGC\,5408: \cite{Kaaretetal2003}; in NGC\,5204: \cite{Liuetal2004}; 
in NGC\,4559: \cite{Soriaetal2005};
in Holmberg IX: \cite{PakullMirioni2002})
the optical counterpart is thought to be
an individual O or B star.

One explanation for the lack of ULX-cluster
associations is that the accreting systems are
runaway binaries ejected from a cluster.
However, the kick velocities required
to explain displacements of a few hundred pc
would rule out BH masses $\gtae 20$ M$_{\odot}$
(\cite{ZezasFabbiano2002}). Hence, the ejection
scenario is not consistent with IMBHs.
An alternative possibility is that the parent 
cluster has already dispersed. Physical mechanisms 
leading to the expansion and disintegration of star 
clusters are tidal disruption, or mass loss 
via SN explosions and winds from evolved 
massive stars (e.g.,  \cite{Hills1980}; \cite{Mathieu1983}).
However, it is doubtful that these processes 
can entirely dissipate a massive 
cluster on timescales as short as $\sim 10^7$ yr.

In this paper, we use simple analytical approximations 
to compare the timescale for IMBH formation 
with characteristic timescales of cluster evolution.
We show that for some density profiles, the collapse 
can already occur during the initial protocluster 
phase, when the stars are still embedded 
in optically-thick gas. We then show that for a range 
of masses and densities, the parent clusters 
may not survive the embedded phase, 
leading to the formation of an apparently isolated IMBH.

\section{VELOCITIES AND DENSITIES IN A CLUSTER CORE}

A useful definition of relaxation timescale for a cluster 
of mass $M$ is, from \cite{Spitzer1987}:
\begin{equation}
t_{\rm r} = \frac{\sigma^3_3}{4\pi (3/2)^{1/2}\,G^2({\rm ln}\,\Lambda) nm^2},
\end{equation}
where $\sigma_3$ is the three-dimensional velocity dispersion, 
$n$ is the number density of stars, $m$ is the average 
stellar mass and ${\rm ln}\,\Lambda$ is the Coulomb logarithm.
For processes related to core collapse, 
we can take the initial values of these quantities 
at the cluster center (e.g., \cite{Gurkanetal2004}); 
we obtain a central relaxation timescale $t_{\rm rc}$:
\begin{equation}
t_{\rm rc}(0) = \frac{3^{3/2}\sigma^3_1(0)}{4\pi (3/2)^{1/2}\,G^2({\rm
ln}\,\Lambda) 
\rho(0) m(0)},
\end{equation}
where $\rho(0)$ is the initial mass density at the core
and $\sigma_1(0)$ is the initial one-dimensional velocity dispersion.
The initial average stellar mass $m(0)$ at the core may be larger 
than the average mass over the whole cluster, 
if there is initial mass segragation. For simplicity, 
we shall assume they are the same.

Assuming for simplicity that the system is virialized, we have 
an additional relation 
between density and velocity dispersion at the cluster 
center. For the Plummer model (\cite{Plummer1915}; \cite{Spitzer1987}), 
\begin{equation}
\sigma^3_3(0) = (1/2)^{3/2}(4\pi/3)^{1/2}\,G^{3/2}M \rho^{1/2}(0), 
\end{equation}
hence
\begin{equation}
\sigma_1(0) = 0.518\,G^{1/2}M^{1/3} \rho^{1/6}(0). 
\end{equation}
Analogous expressions can be derived 
from the virial theorem for the King 
density profiles (\cite{King1966}). For example, 
for the King model with dimensionless central 
potential $W_0=2$ (\cite{BinneyTremaine1987}), 
\begin{equation}
\sigma_1(0) = 0.585\,G^{1/2}M^{1/3} \rho^{1/6}(0),  
\end{equation}
and 
\begin{equation}
\sigma_1(0) = 0.286\,G^{1/2}M^{1/3} \rho^{1/6}(0),  
\end{equation}
for the more centrally concentrated $W_0=9$ profile.

We could approximate the Coulomb 
logarithm $\ln \Lambda \approx 7$--$9$ for the parameter range 
of interest here. More accurately, we use 
the definition of $\Lambda$ (\cite{Spitzer1987}, chapter 2), 
to obtain $\Lambda = 3r_{\rm h}\sigma^2_1/(Gm) \approx 0.4M/m$, 
where $r_{\rm h}$ is the half-mass radius. Expressing 
$M \propto \rho^{-1/2}(0)\sigma^3_1(0)$, where the 
proportionality constant depends on the cluster model, 
we obtain:
\begin{equation}
\Lambda = \alpha G^{-3/2}\rho^{-1/2}(0)\sigma^3_1(0)m^{-1}(0).
\end{equation}
Here, $\alpha = 2.9$, $2.5$, $17.2$ for the Plummer, $W_0=2$ 
and $W_0=9$ King profiles, respectively.
As a first-order approximation, we can also 
substitute $\sigma^3_1(0)m^{-1}(0)$ from Eq.~2, so that:
\begin{equation}
\Lambda \approx \beta t_{\rm rc}(0) \rho^{1/2}(0),
\end{equation}
where $\beta = 0.022$, $0.015$, $0.131$ for the three 
selected models, respectively (in CGS units).

Finally, it can be shown that 
\begin{equation}
r_{\rm h}(0) = \gamma G^{-1/2}\rho^{-1/2}(0)\sigma_1(0),  
\end{equation}
where the constant $\gamma = 1.56$, $1.17$, $13.10$ for the three 
profiles.

Using Eqs.~(2) and (7) one can now plot the central velocity 
dispersion as a function of central density, for a given 
central relaxation timescale: 
$\sigma_1(0) = f[\rho(0); t_{\rm rc}(0)]$.
Using one of Eqs.~(4/5/6), one can plot 
$\sigma_1(0) = g[\rho(0); M]$. From Eq.~9, one can plot 
$\sigma_1(0) = h[\rho(0); r_{\rm h}(0)]$.
See also \cite{Rasioetal2004}, 
in particular their Figure~1.4.

\section{TIMESCALE FOR CLUSTER EVOLUTION}

Numerical simulations (\cite{Gurkanetal2004}) 
for a variety of Plummer and King profiles 
have shown that the timescale for core collapse 
($t_{\rm cc}(0)$) in clusters with a broad mass spectrum 
is proportional to the central relaxation timescale, 
rather than to the relaxation timescale 
at the half-mass radius:
\begin{equation}
t_{\rm cc}(0) \approx 0.15 t_{\rm rc}(0),
\end{equation}
also in agreement with the simulations of 
\cite{PzMc2002}.
The final core mass after the runaway collapse is 
found to be (\cite{Gurkanetal2004}):
\begin{equation}
M_{\rm cc} \approx 0.002 M,
\end{equation}
which is likely to evolve later into a BH 
with mass $M_{\rm BH} \approx 0.001 M$, 
via direct collapse or SN explosion.

In the standard treatment, the timescale in Eq.~(10) 
is compared with the stellar evolution timescale: 
core collapse and the subsequent formation of 
an IMBH are possible only if 
$t_{\rm cc}(0) \ltae 3\ {\rm Myr}$, 
typical lifetime of the most massive stars 
on the main sequence (\cite{PzMc2002}). 
If the core collapse is not completed 
after 3 Myr, it will be stopped by strong mass losses 
from supergiant winds and SNe.
By imposing this constraint, one can identify 
the regions in the $(\sigma_1(0),\,\rho(0))$ plane 
where core collapse is most likely to occur.

The situation becomes more complicated when 
the role of gas in the young clusters is taken 
into account.
Both from an analytical 
approximation and from numerical simulations (e.g., 
\cite{Pz2004}, in particular Figure 1.21)
it appears that centrally-concentrated clusters 
can achieve core collapse in $< 1$ Myr.
If so, the runaway merger 
of the most massive stars occurs when most 
stars in the embedded cluster are still surrounded by 
a spherical cocoon, optically thick 
to the ionizing radiation (class 0/I protostars).
The lifetime of a class I phase 
is $\approx 1$--$3 \times 10^5$ yr 
for low-mass protostars (e.g., \cite{Haischetal2000}; 
\cite{Wilkingetal1989}; \cite{Kenyonetal1990}), 
and probably even longer for O stars, which 
spend $\approx 13$--$15\%$ of their lifetime 
($\approx 3$--$5 \times 10^5$ yr) shrouded 
by an optically thick cocoon
(\cite{KobulnickyJohnson1999}; \cite{WoodChurchwell1989}).
Taking into account that even ``instantaneous'' 
star formation in a young cluster is in fact spread out 
over $\sim$ a few $10^5$ yr, a young cluster 
remains embedded in molecular gas for  
$\approx 0.5$-$1 \times 10^6$ yr (e.g., \cite{Johnson2004}).
At these early ages, the remaining cold gas 
in the young cluster has a mass at least 
comparable to or larger than the mass in stars.

When most of the cocoons dissipate, the cluster gas 
is quickly ionized (on a timescale of $\sim 10^5$ yr) 
by the Lyman continuum photons emitted 
by the OB stars, and reaches a 
characteristic temperature $\approx 10^4$ K. 
For typical densities $\sim 10^3$--$10^6$ cm$^{-3}$, 
this corresponds to pressures $P/k_{\rm B} \sim 10^7$--$10^{10}$ 
K cm$^{-3}$, many orders of magnitude larger than the 
pressure of the ISM. As a consequence, the gas 
in the cluster expands ``explosively'' 
with a velocity of order 
of the sound speed, $v \sim 10$ km s$^{-1}$ 
(\cite{KroupaBoily2002}; \cite{Kroupa2005}). In some young 
clusters in the Antennae, gas expansion velocities 
$\approx 25$--$30$ km s$^{-1}$ have been inferred
(\cite{Whitmoreetal1999}; \cite{Zhangetal2001}).

If the gravitational potential in the cluster 
is too shallow, the expanding gas becomes unbound; 
if the mass loss $\gtae 50\%$, the whole cluster  
will dissipate on the same timescale. 
As a back-of-the-envelope estimate, 
this occurs when the gas expulsion speed 
is larger than the velocity dispersion 
at the half-mass radius: $\sigma_3 (r=r_{\rm h}) \ltae 10$ 
km s$^{-1}$ (\cite{KroupaBoily2002}).
For typical Plummer and King profiles, 
this corresponds to a central velocity dispersion 
$\sigma_1(0) \ltae 15$--$25$ km s$^{-1}$. 
A more accurate calculation of the exact value 
of the velocity dispersion threshold depends 
on the details of the gas and stellar distribution 
and is beyond the scope of this work.
For simplicity, we shall take $\sigma_1(0) < 20$ km s$^{-1}$ as 
a condition for cluster disruption, 
and (conservatively) $t = 5 \times 10^5$ yr for the lifetime 
of an embedded cluster. At the same time, clusters 
have to be more massive than $\sim 10^4 M_{\odot}$ 
for the explosive disruption process to occur: 
smaller clusters do not contain enough O stars 
to ionize all the gas.

The analytical approximation outlined 
in Section 2 did not take into account 
the role of gas. A simple way to account for this effect 
is to substitute the total cluster mass 
$M_{\rm tot} = M + M_{\rm g} \equiv fM$ 
for the stellar mass $M$ in the virial theorem; 
$f$ depends on the star formation efficiency 
and cluster age.
For the young clusters we are dealing with, $f \approx 1.5$--$3$.
As a result, $\sigma_1(0)$ in Eq.~(4/5/6) will be multiplied 
by a factor $f^{1/3}$. For example, taking $f=2$, we shall plot
\begin{equation}
\sigma_1(0) = 0.65\,G^{1/2}M^{1/3} \rho^{1/6}(0)  
\end{equation}
for the Plummer model, and 
\begin{equation}
\sigma_1(0) = 0.36\,G^{1/2}M^{1/3} \rho^{1/6}(0),  
\end{equation}
for the $W_0=9$ King profile, 
where $\rho(0) \equiv fn(0)m(0)$ is now the total initial 
central density. With similar changes we also 
easily modify the other scaling relations in Section 2; 
for example $\Lambda = fM/m \approx (\beta/f) 
t_{\rm rc}(0) \rho^{1/2}(0)$, etc.

\section{IMBHs WITH OR WITHOUT THEIR PARENT CLUSTERS}

Putting together the timescale constraints 
discussed in Section 2 and 3, we identify 
{\it two different regions in the parameter space}, 
relevant to the formation of IMBHs 
from runaway core collapse:
\begin{itemize}
\item
the first regime (region A in Figs.~1--4) 
is the one discussed by \cite{PzMc2002}: core collapse 
on a timescale $t_{\rm cc}(0) \ltae 3$ Myr, 
inside a super star cluster (mass $\sim$ a few $10^5$ 
to a few $10^6 M_{\odot}$), giving rise to a compact remnant 
$\gtae 500 M_{\odot}$. The parent cluster is massive 
enough to survive the initial gas expulsion; it may 
dissipate later, on much longer timescales, 
owing to mass loss in later stages of stellar evolution, 
or tidal interactions. Stellar populations near 
the brightest ULXs in nearby star-forming galaxies 
have typical ages $\sim 10^7$ yr: if the accreting 
compact objects are IMBHs formed through this process, 
we expect them to be still contained inside 
their parent clusters. A likely example is 
the ULX associated to the super star cluster MGG-11 
in M\,82 (\cite{Pzetal2004});
\item the second regime (region B in Figs.~2--4) 
is for $t_{\rm cc}(0) \ltae 0.5$ Myr, 
in clusters that satisfy the disruption condition 
discussed in Section 3: $M \gtae 10^4 M_{\odot}$ 
and $\sigma_1(0) < 20$ km s$^{-1}$. A cluster 
in this region of the parameter space 
is unlikely to survive beyond its embedded phase: its 
stars will keep expanding freely with velocities 
$\sim \sigma_3$. After $10^{7}$ yr, this will result 
in an OB association with a diameter of $\sim 200$ pc, 
difficult to distinguish from other surrounding  
star-forming regions. Not all the clusters in this regime 
are suitable for IMBH formation: assuming that Eq.~(11) holds, 
only the subset of exploding clusters with 
$M \gtae 5 \times 10^4 M_{\odot}$ are massive enough 
to produce BHs with $M \gtae 50 M_{\odot}$, 
required to explain the observed ULX luminosities.
\end{itemize}

The densest young star clusters in our galaxy 
(e.g., RU136, NGC 3606, the Arches) 
have total central densities $\sim 10^6 M_{\odot}$ pc$^{-3}$. 
Therefore, we took $\rho_{\rm max}(0) = 10^7 M_{\odot}$ 
pc$^{-3}$ as a reasonable upper limit to the 
initial central density in our parameter space.

Core collapse is more likely to occur
on short timescales for centrally 
concentrated profiles: for example, for 
the $W_0=9$ King profile (Fig.~2) rather 
than for the Plummer profile (Fig.~1), 
in agreement with the simulations 
of \cite{Pz2004}.
For a given cluster model, the parameter 
space available for runaway core collapse 
is enhanced when the average stellar mass 
is higher, that is when the initial mass 
function is top-heavy. Here we compare 
the initial conditions for $m=0.5 M_{\odot}$ 
(Fig.~3) and $m=3.0 M_{\odot}$ (Fig.~4), 
for the same $W_0=9$ King profile.
For example, values of $m\approx 3.0 M_{\odot}$ 
have been inferred for the MGG-11 cluster in M\,82, 
which is likely to contain an IMBH
(\cite{Pzetal2004}; \cite{McCradyetal2003}).

Taking a $W=9$ King profile with a typical central 
density $\sim 10^6 M_{\odot}$ pc$^{-3}$ (Figs.~2--4),
we infer from our simple analytical approach 
that clusters with initial stellar masses 
$\sim 10^5 M_{\odot}$ may produce a collapsed 
core during their embedded phase, but may not 
survive the ionization/expulsion of their gas. 
The collapsed cores may later evolve into apparently 
isolated IMBHs with masses up to $\approx 200 M_{\odot}$.
On the other hand, super star clusters with initial stellar masses 
$\sim 10^6 M_{\odot}$ 
may survive the embedded phase and produce IMBHs 
with masses up to $\sim 10^3 M_{\odot}$.
Considering that the initial cluster mass function 
${\rm d}N \sim M^{-2}{\rm d}M$, we also expect that 
IMBHs from clusters in region B should be an order 
of magnitude more numerous than those formed 
from clusters in region A. 

\section{CONCLUSIONS}

We have used a simple analytical approximation 
to suggest that runaway core collapse in 
a young star cluster may occur in two 
distinct regimes. The first regime (extensively 
investigated with numerical simulations 
by \cite{Pzetal2004} and 
\cite{Gurkanetal2004}) is a collapse 
on a timescale $\ltae 3$ Myr (main-sequence 
lifetime of the O stars) in a cluster 
of total initial mass $\sim 10^6 M_{\odot}$. Observed 
after $\sim 10^7$ yr, the outcome of the collapse 
is likely to be an accreting IMBH ($M_{\rm BH} 
\sim 10^3 M_{\odot}$) inside a bright 
cluster. The second regime is a collapse 
on a shorter timescale ($\ltae 0.5$ Myr), 
in a smaller cluster (initial mass $\sim 10^5 M_{\odot}$), 
during its initial embedded phase, when 
the most massive protostars are still shrouded 
by optically-thick cocoons. If the velocity 
dispersion in a cluster is smaller than 
the thermal velocity of the ionized gas, 
such a cluster does not survive beyond its embedded 
phase, because of the explosive loss of gas.
Observing the system after $\sim 10^7$ yr, we 
may find an accreting IMBH ($M_{\rm BH} 
\sim 10^2 M_{\odot}$) in a star-forming region 
or OB association, but apparently not 
inside any clusters. 

Observationally, ULXs have been explained 
as accreting IMBHs. Runaway core collapse 
is a possible mechanism of IMBH formation 
in young stellar environments.
However, most ULXs are not coincident with a cluster, 
though they are often associated with OB stars.
We have suggested that the existence of 
two possible regimes of core collapse may 
explain this puzzle. 
The ULX associated with the M\,82 cluster MGG-11 
may be an example of core collapse in the first regime, 
when the parent cluster survives. 
Most of the other ULXs in nearby star-forming galaxies, 
with an OB companion and X-ray 
luminosities up to $\sim 10^{40}$ erg s$^{-1}$, 
may have been formed in the other regime, when 
the parent cluster evaporates explosively.

The parameter space allowing IMBH formation from 
a runaway core collapse is more extended 
for a top-heavy stellar mass function.
It has often been claimed that the stellar 
initial mass function is top-heavy (higher 
fraction of high-mass stars) in some starburst 
galaxies, although the issue is still 
controversial (\cite{Elmegreen2005} and references 
therein). If this is the case, we speculate 
that it may explain why ULXs seem to be more 
often found in starburst galaxies or in environments 
where star formation is triggered 
by galactic mergers or tidal interactions.


Finally, if many ULXs were indeed formed 
in the core of a long-dissolved protocluster, 
we may want to search for clues of that disruption 
event. For example, after $\sim 10^7$ yr, the expanding 
shell or cloud of gas with a mass $\sim 10^5 M_{\odot}$ 
will have a characteristic radius $\sim 100$ pc 
and density $\sim 1$ cm$^{-3}$. This gas 
may now be X-ray photoionized by the accreting 
IMBH. We speculate that at least some of 
the ionized nebulae discovered around nearby 
ULXs (\cite{PakullMirioni2002}) may be related 
to the disruption of the same protocluster 
where the BH was formed.

\begin{figure*}[t]
\centering
\includegraphics[width=90mm, angle=270]{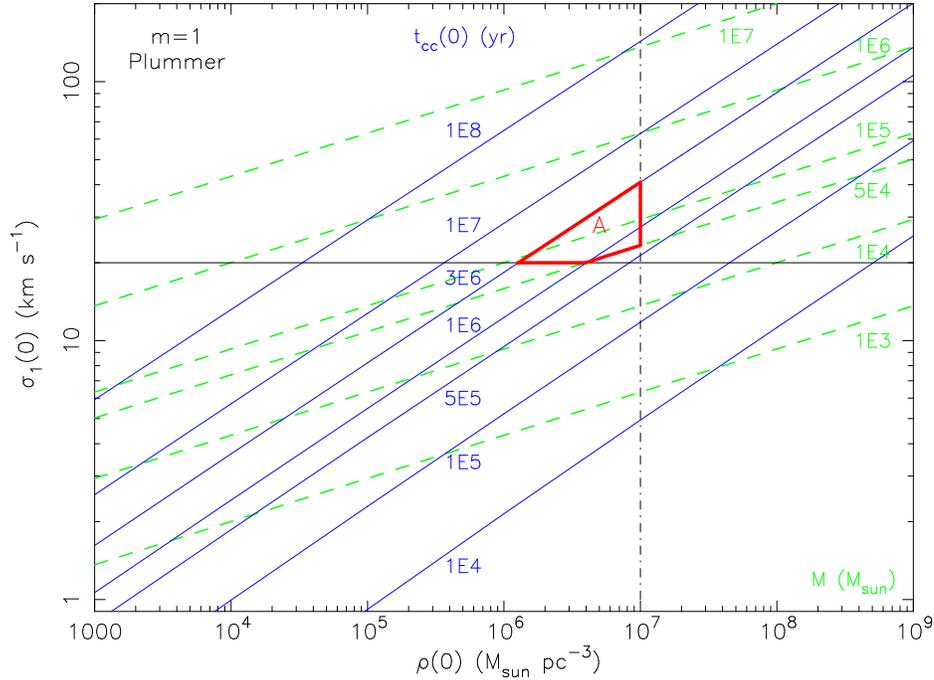}
\caption{Parameter space for the initial central 
density $\rho(0)$ and one-dimensional velocity 
dispersion $\sigma_1(0)$, in a cluster 
with a Plummer profile, with average stellar mass 
$m = 1 M_{\odot}$. We have plotted $\sigma_1(0)$ as a function 
of $\rho(0)$ at fixed core-collapse timescale $t_{\rm cc}(0)$
(solid blue lines), and at fixed total stellar masses $M$ 
(dashed green lines). We have taken 
$t_{\rm cc}(0) = 0.15 t_{\rm rc}(0)$.
A red box (marked with ``A'') identifies the region 
of the parameter space where the runaway core collapse 
occurs in $< 3$ Myr and the cluster is massive enough 
to produce an IMBH ($M_{\rm BH} \approx 10^{-3} M \gtae 50 M_{\odot}$).
Finally, we have divided the parameter space into 
two subsets with $\sigma_1(0) > 20$ and $< 20$ km s$^{-1}$.
Clusters above the threshold are more likely 
to survive their embedded phase; clusters below the threshold 
(but more massive than $\sim 10^4 M_{\odot}$) 
are more likely to evaporate, turning into OB associations.
Finally, we have chosen an initial density 
$\rho(0)=10^7 M_{\odot}$ pc$^{-3}$ as a plausible upper limit, 
based on the values estimated for the densest clusters 
in our and nearby galaxies.} \label{JACpic2-f1}
\end{figure*}

\begin{figure*}[t]
\centering
\includegraphics[width=90mm, angle=270]{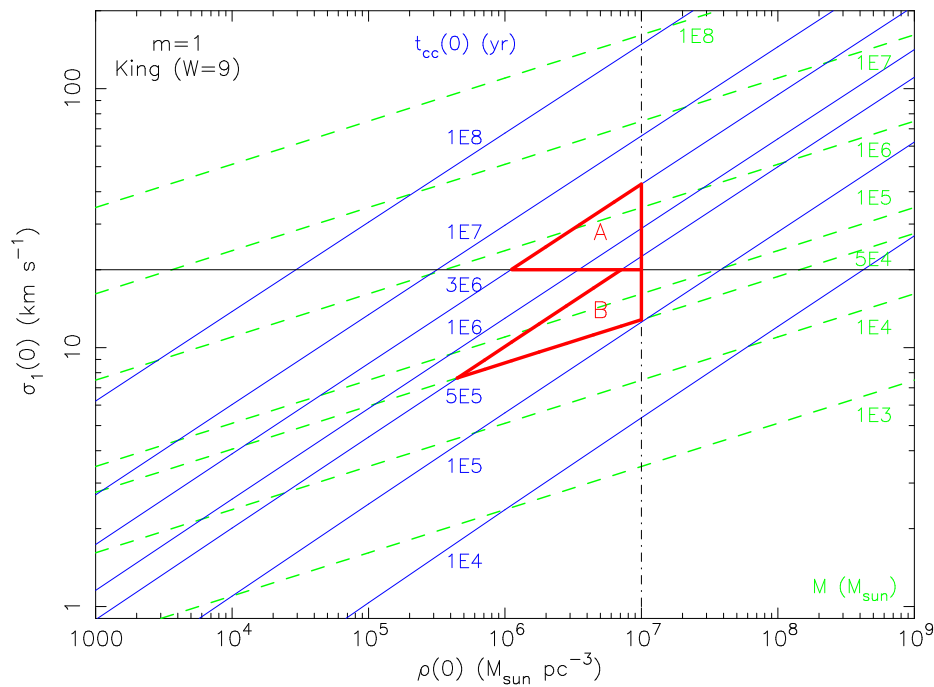}
\caption{As in Figure 1, for a King profile 
of concentration index $W_0=9$. Here there are two distinct 
regions of the initial parameter space where runaway 
core collapse can occur: region A is for massive clusters 
which are expected to survive bound, and region B 
for less massive systems which are expected to evaporate. 
We speculate that IMBHs formed from region-B clusters 
may explain a population of ULXs with X-ray luminosities  
$L_{\rm x} \approx 10^{40}$ erg s$^{-1}$ found 
in nearby star-forming galaxies but not associated 
to any present-day cluster.} \label{JACpic2-f2}
\end{figure*}

\begin{figure*}[t]
\centering
\includegraphics[width=90mm, angle=270]{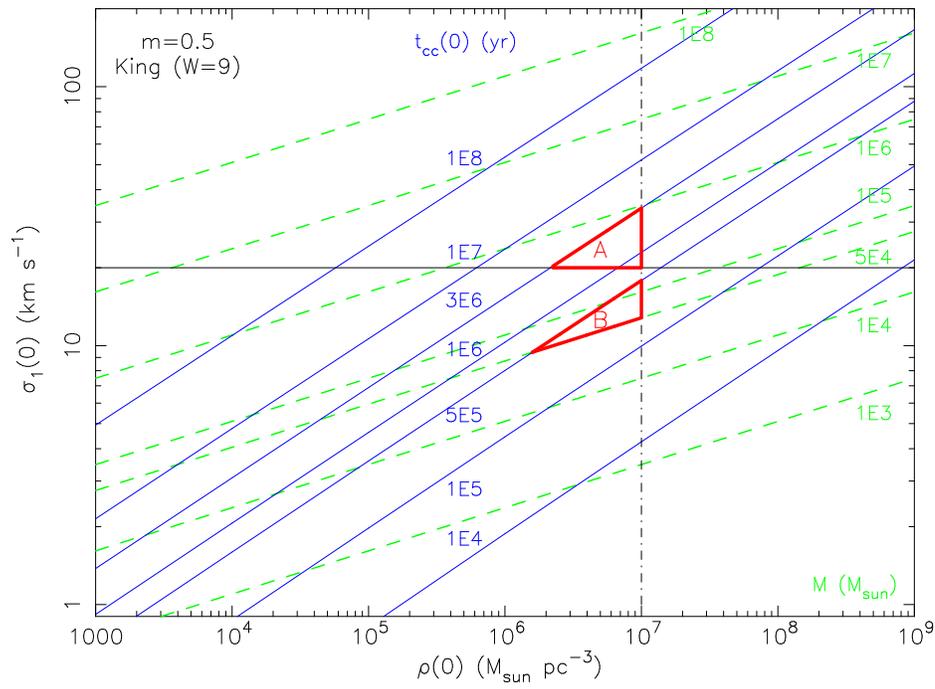}
\caption{As in Figure 2, for an average initial stellar mass 
$m = 0.5 M_{\odot}$. The initial parameter space available 
for runaway core collapse and IMBH formation 
is much reduced.} \label{JACpic2-f3}
\end{figure*}

\begin{figure*}[t]
\centering
\includegraphics[width=90mm, angle=270]{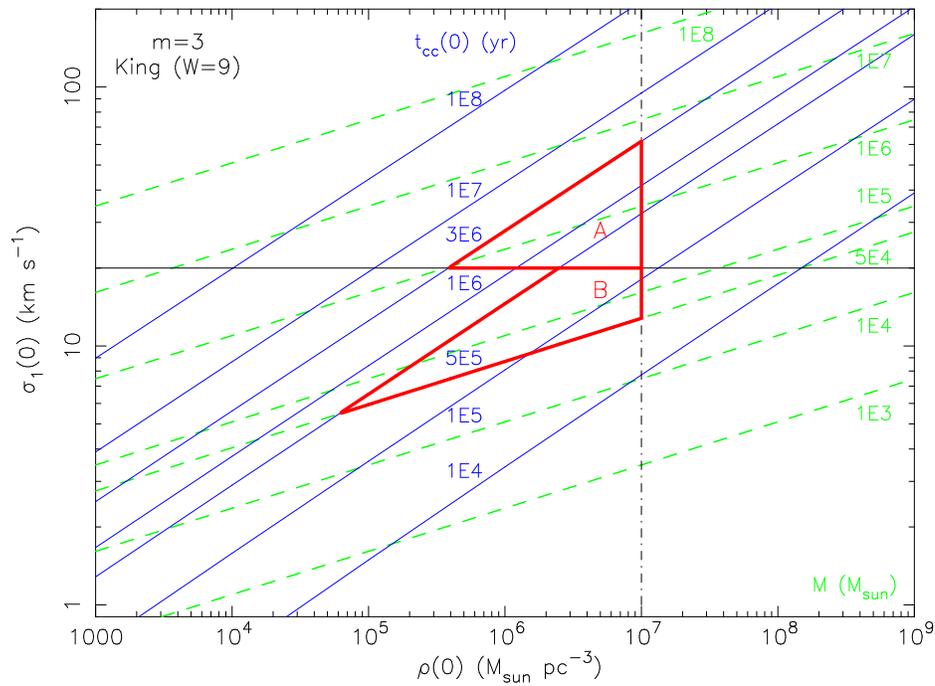}
\caption{As in Figure 2, for an average initial stellar mass 
$m = 3.0 M_{\odot}$. The initial parameter space available 
for runaway core collapse and IMBH formation is 
enhanced for a top-heavy stellar mass function.} \label{JACpic2-f4}
\end{figure*}

\bigskip 
\begin{acknowledgments}
I wish to thank Kenji Bekki, Manfred Pakull and Kinwah Wu for 
useful discussions. My attendance to the Texas Symposium 
was supported by a Royal Society conference grant 
and a ULC graduate school grant.
\end{acknowledgments}

\bigskip 

\begin{thebibliography}{99}   



\bibitem{Begelman}
Begelman, M. C., 2002, ApJ, 568, L97


\bibitem{Kingetal2001}
King, A. R., Davies, M. B., Ward, M. J.,
Fabbiano, G., Elvis, M., 2001, ApJ, 552, L109


\bibitem{FabrikaMescheryakov2001}
Fabrika, S., \& Mescheryakov, A. 2001, 
in the Proceedings of the IAU Symposium 205 (Manchester, August 2000), 
Ed. R. T. Schilizzi, ASP Publication, p. 268 (astro-ph/0103070)


\bibitem{KordingFalckeMarkoff2002}
K\"ording, E., Falcke, H., Markoff, S., 2002, A\&A, 383, L13



\bibitem{ColbertMushotzky1999}
Colbert, E. J. M., Mushotzky, R. F., 1999, ApJ, 519, 89


\bibitem{DavisMushotzky2004}
Davis, D. S., Mushotzky, R. F., 2004, ApJ, 604, 653 

\bibitem{MillerFabianMiller2004}
Miller, J. M., Fabian, A. C., Miller, M. C., 2004, ApJ, 614, L117



\bibitem{MadauRees2001}
Madau, P.,  Rees, M. J., 2001, ApJ, 551, L27


\bibitem{KingDehnen2005}
King, A. R., Dehnen, W., 2005, MNRAS, 357, 275

\bibitem{BekkiFreeman2003}
Bekki, K., Freeman, K. C., 2003, MNRAS, 346, L11



\bibitem{Gurkanetal2004}
G\"urkan, M. A.,  Freitag, M.,
Rasio, F. A., 2004, ApJ, 604, 632

\bibitem{PzMc2002}
Portegies Zwart, S. F.,
McMillan, S., L. W., 2002, ApJ, 576, 899

\bibitem{Pzetal2004}
Portegies Zwart, S. F., Baumgardt, H.,
Hut, P., Makino, J.,
McMillan, S. L. W. 2004, Nature, 428, 724


\bibitem{Spitzer1987}
Spitzer, L., 1987, Dynamical evolution of globular clusters,
Princeton, NJ (Princeton University Press) 

\bibitem{Gaoetal2003}
Gao, Y., Wang, Q., D., Appleton, P. N.,
Lucas, R. A., 2003, ApJ, 596, L171


\bibitem{ZezasFabbiano2002}
Zezas, A., Fabbiano, G. 2002, ApJ, 577, 726

\bibitem{Zezasetal2002}
Zezas, A., Fabbiano, G., Rots, A. H.,
Murray, S. S., 2002, ApJ, 577, 710

\bibitem{SoriaMotch2004}
Soria, R., Motch, C., 2004, A\&A, 422, 915


\bibitem{Smithetal2005}
Smith, B. J., Struck, C., Nowak, M. A. 2005, AJ, 129, 1350


\bibitem{Kaaretetal2003}
Kaaret, P., Corbel, S., Prestwich, A. H., Zezas, A.
2003, Science, 299, 365

\bibitem{Liuetal2004}
Liu, J. F., Bregman, J. N., Seitzer, P. 2004, ApJ, 602, 249

\bibitem{Soriaetal2005}
Soria, R., Cropper, M., Pakull, M.,
Mushotzky, R., Wu, K., 2005, MNRAS, 356, 12

\bibitem{PakullMirioni2002}
Pakull, M. W., Mirioni, L., 2002, (unpublished) proceedings 
of the symposium ``New Visions of the X-ray Universe'', 
ESTEC (The Netherlands), 26-30 Nov 2001; astro-ph/0202488

\bibitem{Hills1980}
Hills, J. G., 1980, ApJ, 235, 986

\bibitem{Mathieu1983}
Mathieu, R. D., 1983, ApJ, 267, L97

\bibitem{Plummer1915}
Plummer, H. C. 1915, MNRAS, 76, 107

\bibitem{King1966}
King, I. R. 1966, AJ, 71, 64

\bibitem{BinneyTremaine1987}
Binney, J., \& Tremaine, S. 1987, Galactic Dynamics 
(Princeton: Princeton Univ. Press)


\bibitem{Rasioetal2004}
Rasio, F. A., Freitag, M., G\"{u}rkan, M. A. 2004, in 
"Carnegie Observatories Astrophysics Series, Vol. 1: 
Coevolution of Black Holes and Galaxies," ed. L. C. Ho 
(Cambridge: Cambridge Univ. Press), p. 138 (astro-ph/0304038)


\bibitem{Pz2004}
Portegies Zwart, S. F. 2004, in "Joint Evolution of Black Holes 
and Galaxies", IOP Publishing (Bristol and Philadelphia, 2005), 
eds M. Colpi, V.Gorini, F.Haardt and U.Moschella (astro-ph/0406550)

\bibitem{Haischetal2000}
Haisch, K. E. Jr., Lada, E. A., Lada, C. J. 2000, AJ, 120, 1396

\bibitem{Wilkingetal1989}
Wilking, B. A., Lada, C. J., Young, E. T. 1989, ApJ, 340, 823

\bibitem{Kenyonetal1990}
Kenyon, S. J., Hartmann, L. W., Strom, K. M., Strom, S. E. 1990, AJ, 
99, 869

\bibitem{KobulnickyJohnson1999}
Kobulnicky, H. A., Johnson, K. E. 1999, ApJ, 527, 154

\bibitem{WoodChurchwell1989}
Wood, D. O., Churchwell, E. 1989, ApJ, 340, 265

\bibitem{Johnson2004}
Johnson, K. E. 2004, in "The Formation and Evolution 
of Massive Young Star Clusters," eds. H. Lamers, A. Nota 
and L. Smith (San Francisco: ASP), astro-ph/0405125


\bibitem{KroupaBoily2002}
Kroupa, P., Boily, C. M., 2002, MNRAS, 336, 1188

\bibitem{Kroupa2005}
Kroupa, P. 2005, to appear in the Proceedings of the Symposium 
"The Three-Dimensional Universe with Gaia", 4-7 October 2004, 
Observatoire de Paris-Meudon, France, eds: C. Turon, 
K.S. O'Flaherty, M.A.C. Perryman (ESA SP-576)

\bibitem{Whitmoreetal1999}
Whitmore, B. C., Zhang, Q., Leitherer, C., Fall, S. M., 
Schweizer, F., Miller, B. W. 1999, AJ, 118, 1551

\bibitem{Zhangetal2001}
Zhang, Q., Fall, S. M., Whitmore, B. C. 2001, ApJ, 561, 727

\bibitem{McCradyetal2003}
McCrady, N., Gilbert, A. M., Graham, J. R. 2003, ApJ, 596, 240

\bibitem{Elmegreen2005}
Elmegreen, B. G. 2005,
to appear in the proceedings of ``Starbursts: from 30 Doradus 
to Lyman Break Galaxies'' Institute of Astronomy, Cambridge
University (September 2004), Kluwer Academic Publishers, 
eds. R. de Grijs and R. M. Gonzalez Delgado (astro-ph/0411193)











\end{thebibliography}

\end{document}